\documentclass[twocolumn]{aastex631}

\usepackage{times}
\usepackage{amsmath, amssymb}
\usepackage{url}

\usepackage[]{xcolor}
\definecolor{myg}{cmyk}{0.75002,0,1,0}

\definecolor{msnote}{hsb:rgb}{0.492,0.492,0.492}

\begin{document}

\title{The delayed radio emission in the black hole X-ray binary MAXI J1348$-$630}

\correspondingauthor{Bei You, Shuai-Kang Yang}
\email{youbei@whu.edu.cn, yangsk@whu.edu.cn}

\author[0000-0002-8231-063X]{Bei You}
\affiliation{School of Physics and Technology, Wuhan University, Wuhan 430072, People’s Republic of China}

\author{Shuai-kang Yang}
\affiliation{School of Physics and Technology, Wuhan University, Wuhan 430072, People’s Republic of China}

\author[0000-0002-5385-9586]{Zhen Yan}
\affiliation{Shanghai Astronomical Observatory, Chinese Academy of Sciences (CAS), Shanghai 200030, China}

\author{Xinwu Cao}
\affiliation{Institute for Astronomy, School of Physics, Zhejiang University, 866 Yuhangtang Road, Hangzhou, 310058, People’s Republic of China}

\author{Andrzej A. Zdziarski}
\affiliation{Nicolaus Copernicus Astronomical Center, Polish Academy of Sciences, Bartycka 18, PL-00-716 Warszawa, Poland}

\begin{abstract}

We explore the coupling between the accretion flow and the jet in black hole X-ray binary (BHXRB) MAXI J1348-630 by analyzing the X-ray and radio observations during its 2019 outburst. We measure the time delay between the radio and Comptonization fluxes with the interpolated cross-correlation function. For the first time, we find that the radio emission lags behind the X-ray Comptonization emission by about 3 days during the rising phase covering the rising hard state and the following soft state. Such a long radio delay indicates that the Comptonization emission most likely originates from the advection-dominated accretion flow rather than the jet in this source. 
The Comptonization luminosity $L_{\rm C}$ in 0.1-100 keV and the radio luminosity $L_{\rm R}$ at 5.5 GHz, after considering the radio delay of $\sim 3$ days, follow the correlation with a slope $\beta = 3.04 \pm 0.93$, which is much steeper than the previously reported $\beta = 0.6$ or 1.40 using the total luminosity in the limited band (e.g., 1-10 keV) in the literature. This highlights the necessity of considering (1) the time delay, (2) the spectral decomposition, and (3) the broad energy band, in the radio-X-ray correlation analysis.
As the jet reappears during the decaying phase (covering the soft state and the following decaying hard state) and the mini-outburst, the Componization and the radio emission appear to be almost simultaneous. And, the radio-Compton correlation during the mini-outburst becomes shallow with the correlation slope $\beta = 1.11 \pm 0.15$. These indicate an intrinsic difference in the accretion-jet coupling physics between the main outburst and the mini-outburst.
\end{abstract}


\section{Introduction} \label{sec:intro}

For a black hole X-ray binary (BHXRB) in an outburst, the observed X-ray radiation originates from the accretion flow near a BH by viscously releasing its gravitational energy \citep{ss1973}. The observed radio emission and high angular resolution radio image indicate 
a relativistic jet launched near the BH \citep{fender2004araa}. The magnetic field is essential for the jet formation via Blandford \& Znajek (BZ) and/or Blandford \& Payne (BP) mechanisms \citep{blandford1977,blandford1982}. 
The accretion process and the related accumulation of the magnetic field near the BH have always been important topics in BH astrophysics. \citep{tchekhovskoy2011}.

Multi-wavelength observations, from radio to X-rays and gamma-rays, are essential to understand the physical process of the magnetized flow accreted onto a BH \citep{jain2001,homan2005,russell2006,dincer2012,kalemci2013,zdziarski2016,weng2021,thomas2022,you2023}. 
In the past, the broadband spectral energy distribution fitting \citep{marino2021,rodi2021,ozbey2022,banerjee2024,et2024,yoshitake2022} and the correlations between the radio/X-ray \citep{corbel2003,coriat2011,williams2022}, are widely used to study the origin of the multi-wavelength emission and the relevant physical process. The monochromatic luminosities at 5 GHz and the integrated 1–10 keV X-rays luminosity and/or the bolometric luminosity are usually used in the studies of the luminosity correlations \citep{carotenuto2021/505}. Moreover, the evolution of the spectral components (via the decomposition in the fits) with the time was considered in the past studies \citep{you2021,fijma2022,dai2023,peng2023}. In recent years, many multi-wavelength spectral timing studies on BHXBs have been conducted to investigate the broad-band variability originating from jets \citep{Tetarenko2019,Tetarenko2021b,Zdziarski2022APJ}. These studies have identified correlations between the X-ray and radio emissions, as well as time lags of tens of minutes between the optical, X-ray, and radio bands, which provide constraints on the jet composition \citep{Tetarenko2021b}.

\cite{you2023} studied the multi-wavelength observations of MAXI J1820+070 in the 2018 outburst. It was found that the radio emission lags behind the X-ray Comptonization emission by about 8 days in the decaying hard state, which is the longest in known low-mass BHXRBs. 
The radio emission indicates the launching and suppression of a jet, as the magnetic fields in the disk are dragged into the hard X-ray emitting advection-dominated accretion flow (ADAF) near the BH \citep{cao2011,yuan2014,2023ApJ...944..182D}.
Such an unprecedented delay of the radio emission was attributed to the expanding of the ADAF (i.e., the thin disk is receding away from the BH, in the truncated disk model). It was demonstrated in \cite{you2023} that, the magnetic field is transported and continuously amplified by the expanding ADAF, even after the hard X-ray peak. Moreover, at the radio peak, i.e., lagging the X-ray peak by $\sim$ 8 days, the accumulated magnetic field eventually becomes dynamically important (i.e., the magnetic force comparable to gravity forces) in the inner edge \citep{narayan2003}. For the first time, how a MAD is formed in a BH accretion system is revealed in \cite{you2023}.

It is then unclear if such a long daily delay between the radio emission and the hard X-ray emission discovered in MAXI J1820+070 exists in other BHXRBs. Furthermore, in some BHXRBs the main outbursts are followed by a few subsequent mini-outbursts before they evolve back to quiescence \citep{mt2017,yan2017, zhanggb2019}. So it is also interesting to study if the radio/X-ray delay behavior and the associated physical process proposed for the main outburst apply to the mini-outburst.

MAXI J1348$-$630 is a recently discovered BHXRB with high-cadence monitoring in radio and X-ray, during the outburst including the rising/decaying hard state and mini-outburst \citep{jana2020, tominaga2020, zhang2020,carotenuto2021/504, garcia2021, zhangw2022}. These observations make MAXI J1348$-$630 an interesting source that enables unprecedented insight into the accretion process of the flow and the magnetic field near the BH. In this work, we will study the time delays between the radio emission and the X-ray Comptonization of MAXI J1348–630 in the main outburst and the mini-outburst.

\section{Data} \label{sec:data}

\begin{figure*}
\centering
\includegraphics[width=\textwidth]{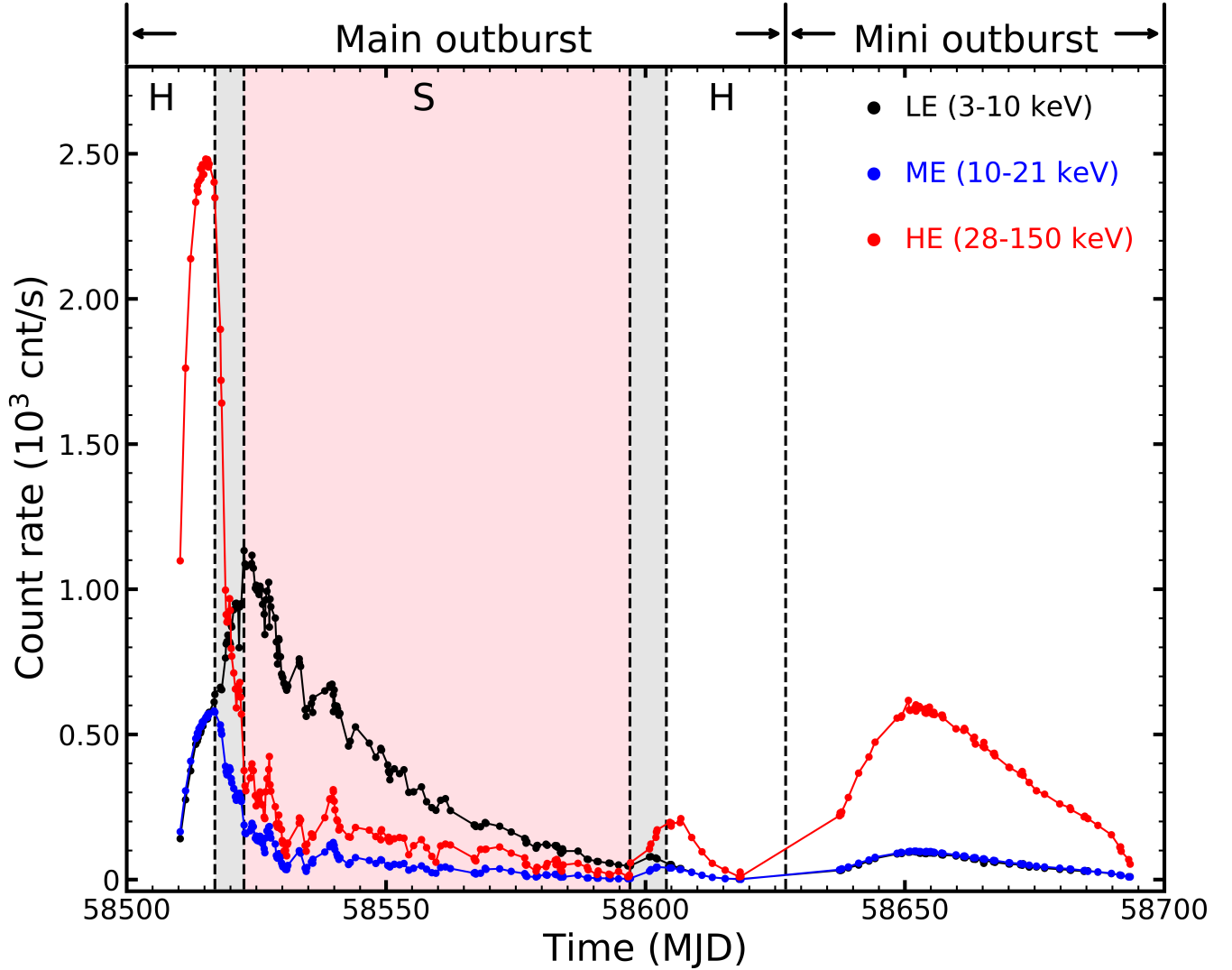}
\caption{
\emph{Insight}-HXMT lightcurves of MAXI J1348–630 in the HE (28-150 keV, red
curve), ME (10-21 keV, blue curve), and LE (3-10 keV, black curve) bands. The two white regions indicate the hard states. The two light grey shaded regions indicate the intermediate states, and the light pink region represents the soft state. The light green shaded region indicates the hard-state-only mini-outburst. Vertical solid lines separate these epochs. Beyond MJD 58600, the count rate of LE is close to the count rate of ME, so it is hard to distinguish the two lightcurves in the figure visually.
}
\label{lightcurve}
\end{figure*}

In X-ray, the long-term and high-cadence observation of MAXI J1348$-$630 by the \emph{Insight}-HXMT was carried from 2019-01-27 (MJD 58510) to 2019-07-29 (MJD 58693). 
Figure \ref{lightcurve} shows the \emph{Insight}-HXMT lightcurves, displaying the main outburst (from MJD 58510 to 58618) and the first mini-outburst (from MJD 58635 to 58700). 
The count rates detected by LE/ME/HE instruments display two apparent flares during the main outburst, and one flare during the mini-outburst.
According to the spectral and timing analysis of the Neutron Star Interior Composition Explorer (NICER) X-ray data \citep{zhang2020}, the rising hard state of the main outburst starts from MJD 58510 to 58517. And, the hard-to-soft state transition ranges from MJD 58517 to 58522.6. Then the source remains in the soft state from MJD 58522.6 to 58597. The soft-to-hard state transition and the decaying hard state, accompanied by a hard X-ray emission recovery, correspond to MJD 58597 to 58620. In the mini-outburst, the source is always in the hard state.


In radio, MAXI J1348$-$630 was monitored by MeerKAT at 1.28 GHz and the Australia Telescope Compact Array (ATCA) at 5.5 and 9 GHz, covering the main outburst and the mini-outburst, which were presented in \cite{carotenuto2021/504}. 
The radio spectral index $\alpha$ ($S_{\nu} \propto \nu^{\alpha}$) for each observation of MeerKAT, is estimated by interpolating the measured index at the adjacent observations by ATCA. Then, 
we convert the MeerKAT radio flux density at 1.28 GHz to the 5.5 GHz monochromatic flux with the interpolated spectral index. 
The evolution of the estimated radio fluxes at 5.5 GHz for the main outburst and mini-outburst are plotted in Fig. \ref{multi_lc}.



\section{X-ray spectral analysis} \label{sec:spectral fits}

We perform the spectral fits to the observations of LE/ME/HE instruments onboard \emph{Insight}-HXMT in the 2--150 keV energy range with {\tt XSPEC}. The data in the 21--24 keV energy range is ignored in the spectral fits, due to the photoelectric effect of electrons in Silver K-shell in the ME detectors.

We first fit the spectra with a simple model {\tt Tbabs*(diskbb+nthcomp)}. {\tt TBabs} \citep{wilms2000} accounts for the Galactic absorption, where the column density is fixed to $N_\mathrm{H} = 8.6\times 10^{21} \rm cm^{-2}$ \citep{tominaga2020}. The {\tt diskbb} represents the thermal emission from the accretion disk \citep{mitsuda1984}. The model {\tt nthcomp} represents the Comptonization by the hot electrons \citep{Zdzi1996}.
The residuals reveal the line features at about 3-10 keV and the hump above 20 keV, which indicate the disk reflection \citep{garcia2013}. Thus, we also try the model {\tt Tbabs*(diskbb+relxillCp)} in the spectral fits, where {\tt relxillCp} includes the Comptonization and the relativistic reflection. 
It turns out that the spectra are fitted well with the reduced $\chi^2 \lesssim 1.1$.
Note that the spectra during the mini-outburst can be fitted with the model {\tt Tbabs*relxillCp} without the requirement of the disk component, since the disk component is too weak to be detected by \emph{Insight}-HXMT. As for the model configuration, the inner radius of the accretion disk in {\tt relxillCp} is tied to the normalization of {\tt diskbb}, using the formula $Norm = (R_{in}/D_{10})^{2}\cos{\theta}$, where $D_{10}$ is the distance to the source in units of 10 kpc, and $\theta$ is the inclination angle. Here we use a mid-to-low inclination $\theta = 29.3^{\circ}$, supported by the orientation of the jet \citep{carotenuto2022a}.

We note that the spectra cannot be fitted well at the end of the rising phase (after MJD 58536). The reduced $\chi^{2}$ are relatively large with the reduced $\chi^2 > 1.4$, and the residuals show a bump feature at about 5 keV. In order to find out the reason, we compare the \emph{Insight}-HXMT spectral data with the NICER spectral data, by fitting the spectra with the same model. 
It turns out that there are unexpected features around 5 keV in the ratio plots for the \emph{Insight}-HXMT, compared to the NICER observations. Therefore, the following analysis does not include the \emph{Insight}-HXMT observations after MJD 58536 within rising phase.



\begin{figure*}
\centering
\includegraphics[width=\textwidth]{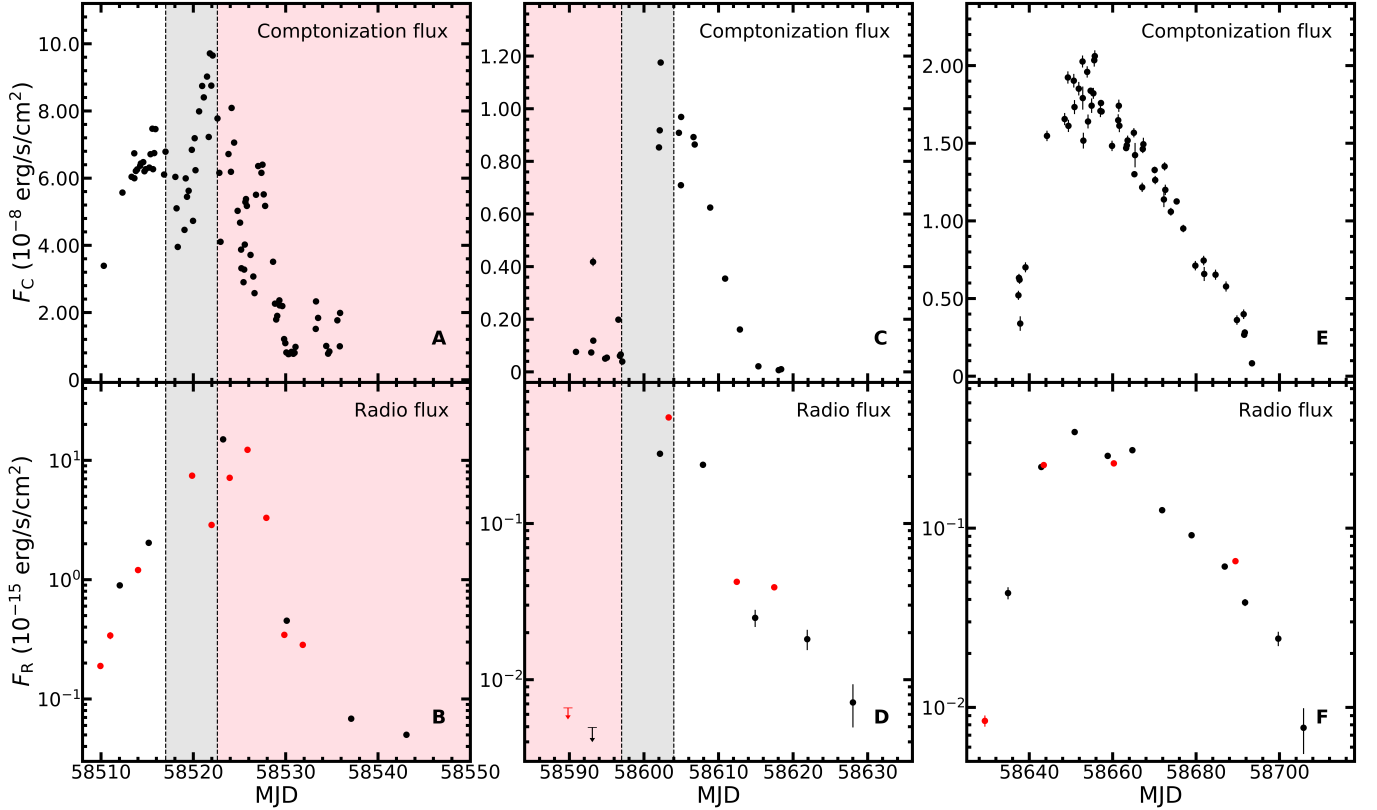}
\caption{
The evolution of the fluxes with time during the rising phase (panel A-B), the decaying phase (panel C-D) and the mini-outburst (panel E-F). Panel A: The estimated Comptonization flux in the 0.1-100 keV band during the rising phase. Panel B: The radio flux at 5.5 GHz by \emph{ATCA} (red points) and \emph{MeerKAT} (black points) during the rising phase. Panel C: The estimated Comptonization flux in the 0.1-100 keV band during the decaying phase. Panel D: The radio flux at 5.5 GHz by \emph{ATCA} (red points) and \emph{MeerKAT} (black points) during the decaying phase. Panel E: The estimated Comptonization flux in the 0.1-100 keV band during the mini-outburst. Panel F: The radio flux at 5.5 GHz by \emph{ATCA} (red points) and \emph{MeerKAT} (black points) during the mini-outburst. The radio emission was detected from the MAXI J1348-630 core location. The MeerKAT radio flux density at 1.28 GHz is converted to the 5.5 GHz monochromatic flux with the interpolated spectral index from the ATCA observations. The white, grey and pink region represent the hard,intermediate and soft states,respectively. 
}
\label{multi_lc}
\end{figure*}

Based on the spectral fits, we use the {\tt cflux} in {\tt XSPEC} to estimate the X-ray Comptonization flux $F_{\rm C}$ in 0.1-100 keV. The evolution of the estimated X-ray flux for the main outburst and mini-outburst are plotted in Fig. \ref{multi_lc}. 

\section{Time-delay between the X-ray and radio emission}
\begin{figure}
\centering
\includegraphics[width=0.46\textwidth]{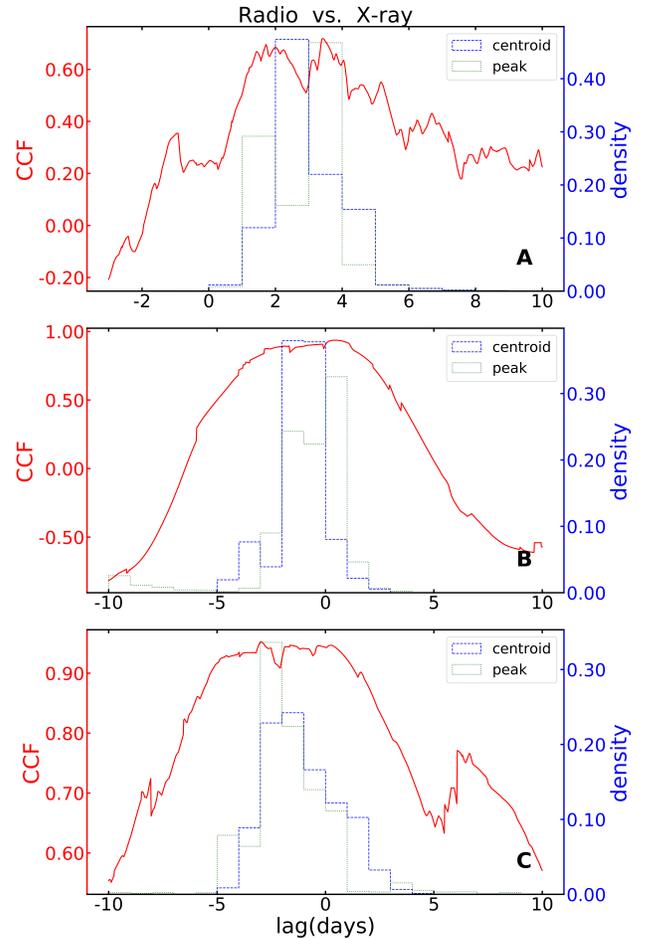}
\caption{
Panel A: Cross-correlation analysis between the radio and the X-ray fluxes during rising phase. Panel B: Cross-correlation analysis between the radio and the X-ray fluxes during decaying phase. Panel C: Cross-correlation analysis between the radio and the X-ray fluxes during mini-outburst. In all panels, the dashed and dotted histograms in blue are the cross-correlation centroid lag distribution and peak lag distribution obtained by the flux randomization/random subset sampling (FR/RSS) Methods, respectively, the axis of which is provided on the right side. 
}
\label{rx_ccf}
\end{figure}

In this work, we aim to study the coupling between the accretion flow and the jet by measuring the time lags between the radio/hard X-ray fluxes, with the interpolated cross-correlation function \citep[ICCF;][]{kaspi2000,you2023}. More details of the implementation of ICCF for BHXRBs can be found in \cite{you2023}. Accompanying the three hard X-ray flares (see HE in Fig. \ref{lightcurve}), the radio emission also displays three flares (see Fig. \ref{multi_lc}). 
We refer to the period of MJD 58510 -- 58543 as the rising phase, covering the rising hard state, the hard-to-soft state transition, and the early-time soft state (see panel A-B in Fig. \ref{multi_lc}), refer to the period of MJD 58590 -- 58628 as the decaying phase, covering the late-time soft state, the soft-to-hard state transition, and the decaying hard state (see panel C-D in Fig. \ref{multi_lc}), and refer to the period of MJD 58620--58710 as the mini-outburst (see panel E-F in Fig. \ref{multi_lc}). The three hard X-ray flares are dominated by the Compton emission, we then analyse the ICCF between the Compton and radio fluxes during the above three flares (see \autoref{rx_ccf}).

During the rising phase, it is evident that there is an X-ray Comptonization emission and a radio emission (see panel A-B in Fig. \ref{multi_lc}).
Between MJD 58510 and 58522, the radio spectral indices $\alpha > 0$ indicate that the observed radio emission from the core location is likely to be associated with the compact jets during this period \citep{carotenuto2021/505}, given the flat or slightly inverted spectrum as a signature of self-absorbed synchrotron emission from a compact radio jet \citep{markoff2001,corbel2002,russell2022}. However, as it was transitioning to the soft state (after around MJD 58522.6), the radio emission becomes a steep spectrum with spectral index $\alpha < 0$ \citep{carotenuto2021/505}. In this case, it is hard to determine whether the compact jet dominates the radio flux, as the discrete ejection likely contributes significantly to the observed radio emission \citep{carotenuto2021/505}. 
Nontheless, the ICCF analysis between these two lightcurves in Fig. \ref{rx_ccf}A shows that the radio emission lags the X-ray emission with the peak time delay of $\Delta \tau_{\rm rx}^{\rm p} = 3.25^{+0.53}_{-1.55}$ days and the centroid delay of $\Delta \tau_{\rm rx}^{\rm c} = 2.82^{+1.22}_{-0.73}$ days. 
We also repeat the ICCF analysis but use the lightcurves between MJD 57510 and 58522.6 when the radio fluxes are believed to come from the compact jet. It turns out that the radio emission still lags behind the X-ray emission by about 3 days.  


Panel C-D in Fig. \ref{multi_lc} show the X-ray/radio flares during the decaying phase. As the disk flux continuously decays with time, the X-ray Comptonization and radio emissions are observed during the decaying phase. The ICCF analysis indicates that the radio emission lags behind the X-ray emission by the peak time delay of $\Delta \tau_{\rm rx}^{\rm p} = -0.74^{+1.40}_{-1.24}$ days and the centroid delay of $\Delta \tau_{\rm rx}^{\rm c} = -1.02^{+0.76}_{-0.87}$ days. However, given the relatively sparse sampling of the radio measurements and the uncertainties in the derived delays, there is most likely no daily delay between the X-ray Comptonization and the radio emission, i.e., nearly simultaneous hard X-ray/radio emission. This is quite different from the above results for the rising phase during which the radio lags behind the X-ray Comptonization by about 3 days. And is also opposite to the reported radio delay in the decaying hard state of MAXI J1820+070 \citep{you2023}.

As for the mini-outburst, panel E-F in Fig. \ref{multi_lc} show the evolution of the X-ray Comptonization and radio fluxes over time. The ICCF analysis indicates the radio emission lags the X-ray emission by the peak time delay $\Delta \tau_{\rm rx}^{\rm p} = -1.98^{+1.88}_{-1.05}$ days and the centroid delay $\Delta \tau_{\rm rx}^{\rm c} = -1.42^{+2.26}_{-1.26}$ days. Given the relatively sparse sampling of the radio measurements and the uncertainties in the derived delays, there is most likely no daily delay between the X-ray Comptonization and the radio emission during the mini-outburst, i.e., nearly simultaneous hard X-ray/radio emission, which is similar to the results for the decaying phase during the main outburst.



\section{Discussion}

The observed radio emission originates from the jets, depending on the magnetic field accumulated by the accretion \citep{blandford1979}. However, the origin of the hard X-ray Componization is uncertain, which may be the jet base or the ADAF \citep{yuan2014}. The time delay between the radio emission and the Comptonization could be used to study the origin of the X-ray emission and the associated accretion process. The observed 8-day radio lag relative to the hard X-ray emission in MAXI J1820+070 indicates that the X-ray Comptonization during the decaying hard state in its 2018 outburst originates from the ADAF, rather than the jet \citep{you2023}.

In previous sections, we studied the time delays between the radio emission of the jet and the X-ray Comptonization of MAXI J1348–630 in the main outburst and the hard-state-only mini-outburst. 
In the following, we will use the discovered time delays between the radio emission of the jet and the X-ray Comptonization, to discuss the time evolution of the accretion flow and the magnetic field during the outburst of this source. 

\subsection{Delayed radio emission during the rising phase}

As shown in Fig. \ref{multi_lc}, during the rising phase, the onsets of both the radio emission and X-ray Comptonization were detected around MJD 58510. After that, the X-ray Comptonization first peaks around MJD 58516, and then fades towards the hard-to-soft state transition, showing a typical X-ray Compton variation in the rising hard state \citep{you2021}. Surprisingly, after the X-ray Compton peaking, the radio emission continuously increases with time, and appears to peak around MJD 58520. The ICCF analysis reveals that the radio emission lags the X-ray Comptonization emission by about 3 days. Considering the scenario that the radio delay is caused by jet traveling, we estimate the time lag to be the order of tens of minutes \citep{you2023}. This is significantly shorter than the observed radio lag of $\sim$ 3 days. Thus, the daily optical/radio delays in this work cannot be attributed to the jet traveling, and we rule out the possibility that X-rays are emitted by the jet in MAXI J348-630. The discovered radio delay suggests that the Compton emission originates from the ADAF rather than the jet during the rising phase in this source, and the Compton luminosity light curve indicates the evolution (the building up and fading away) of the ADAF. 



During the state transition, the ADAF is accreting at a rate very close to the critical rate, and the radial energy advection is negligible \citep{1995ApJ...452..710N}. Thus, almost all dissipated gravitational power in the ADAF can be radiated out. The flux of the inner ADAF is $F_{\rm ADAF}\propto \dot{m}(1/r_{\rm in}-1/r_{\rm tr})$, where $r_{\rm tr}$ is the truncated radius between inner ADAF and the outer thin disk. In the hard-to-soft state transition, the mass accretion rate increases with decreasing truncated radius \citep{1997ApJ...489..865E}. This implies that the ADAF radiates at its peak flux $F_{\rm ADAF}^{\rm peak}$ while the increase of the accretion rate is balanced by the decline of the truncated radius $r_{\rm tr}$. We note that the radio emission from the jets in this source is rather strong while it radiates at the peak flux in the hard X-ray band. However, the radio flux continuously rises till its peak at about 3 days after the X-ray peak. We conjecture that the large-scale magnetic field has already been sufficiently strong to arrest the accretion flow while the inner ADAF is radiating at its peak flux \citep{2003ApJ...592.1042I,narayan2003,2022MNRAS.511.2040B}. The magnetic field of the MAD can be estimated as $B_{\rm MAD}\propto \dot{m}^{1/2}$ \citep[see Narayan et al. 2003, and Equation S8 in][]{you2023}. The accretion rate keeps on increasing after the hard X-ray peak, and then the field in the MAD increases accordingly. The radio flux increases with the field strength of the MAD if the jets are assumed to be launched via the Blandford-Znajek mechanism \citep{blandford1977}. The truncated radius is approaching the ISCO with an increasing accretion rate, which leads to less amplification of the field in the shrinking ADAF \citep[][]{cao2011}. Ultimately, the field in the inner edge of the ADAF is too weak to meet the MAD criterion, and therefore, the radio emission from the jets declines afterward. 

The production of transient ejections has been observed in a number of sources during short excursions from the soft to the intermediate state, and back to the soft state, when crossing the `jet line' \citep[e.g.][]{brocksopp2013}. During the hard state, the compact jets are rarely resolved in radio, with their extent $\lesssim 10^{15} \, \rm cm$. However, during the hard-to-soft state transition, the transient jets could propagate far away from the core, reaching the distances of $\sim 10^{18} \, \rm cm$, which are at least three orders of magnitudes further away than the compact jet \citep[see][and the references therein]{aaz2024}. The power of the compact jet in the hard state, if assuming electron-positron pair dominance, is less than the MAD limit \citep{aaz2024}, which is consistent with the argument of a low magnetic field in a recent study of X-ray polarization \citep{barnier2024}. This implies that the compact jet in the bright hard state is launched from the standard and normal evolution (SANE) accretion flow \citep{narayan2012}, which agrees with the argument of \cite{fragile2023}. It was then proposed in \cite{aaz2024} that the difference in the propagation length between the compact jet during the bright hard state and the transient jet during the state transient could be attributed to both the jet composition and the power. Note that, in the case of the faint hard state, e.g., when a BHXRB leaves the soft state, the accretion mass was low compared to the bright hard state, and it was discovered to decay exponentially with time. In this case, the ADAF was observationally inferred to expand. In this case, the criteria for reaching the MAD state decreases with time \citep{narayan2003}, whereas the magnetic field amplification becomes significant \citep{you2023}. It was proposed that the inner region of the ADAF would become magnetically dominated, forming the MAD in the decaying hard state.

As for MAXI J1348-630 studied in this work, the jet power during the 
hard-to-soft state transition was estimated to be at the MAD limit \citep{aaz2023}, in which case the ejection was observed up to almost a parsec \citep{carotenuto2021/504}, similar to MAXI J1820+070 \citep{aaz2024}. In contrast, the compact jets in this source were not resolved and
certainly did not propagate to large distances. Similar to MAXI J1820+070 discussed above, such different propagation lengths of the compact and the transient jets, may imply that the compact jet in the bright hard state is launched from the SANE while the transient jet during the state transient is associated with the MAD.

We also note that, as the radio emission decreases during the soft state, the optically thin radio emission from the core location was again detected between MJD 58573 and 58582, with the flux density increasing by at least two orders of magnitude \citep{carotenuto2021/504}. Interestingly, during such a short radio flare, the X-ray flux changed by less than about 30\%. The physics mechanism behind these radio and X-ray activities in the soft state is interesting but unclear, which should be paid attention to in future studies of jet physics.

It is well known that the jet behavior correlates with the X-ray spectral states \citep{fender2004}. The jet quenching is usually associated with the hard-to-soft state transition, while the hard X-ray from the ADAF/corona decreases and the soft X-ray from the disk still increases \citep{Fender1999}. Current observations show that the relation between the jet quenching and the decreasing of hard X-rays is complex. The nearly simultaneous hard X-ray/radio emission during the outburst rise of MAXI J1820+070 indicates that the jet quenching was almost simultaneous with the weakening of the ADAF/corona \citep{you2023}, which was also observed in the 1998 outburst of GX 339-4 \citep{Fender1999}. However, the radio-X-ray time lag in MAXI J1348-630 indicates that the jet started to be quenched at about 3 days after the state transition, which is similar to the 1999 outburst of XTE J1859+226 \citep{Brocksopp2002}. 
\citet{Yan2012} even discovered that the jet quenching started 10 days before the state transition. \cite{Russell2020} reported an onset of jet quenching in the infrared band in MAXI J1535-571, suggesting that the jet quenching may start earlier than predicted by radio detection. To better understand the coupling between the jet and hot corona/ADAF, more high-cadence multi-wavelength observations across the state transition are required in the future.



\subsection{Quasi-simultaneous radio emission during decaying phase and the mini-outburst}

The ICCF analysis shows that the X-ray Comptonization and radio synchrotron emission appear to be nearly simultaneous. 
The simultaneous hard X-ray/radio emissions could be attributed to the possibility that the X-ray emission originates from the Comptonization within the radio jet \citep{markoff2005}. Alternatively, the X-ray emission may come from the ADAF, rather than the radio jet \citep{dai2023, you2023}. In this case, the accretion flow and the magnetic field threading it in the outer thin disk are efficiently dragged to pass through the ADAF with large radial velocity, towards the BH \citep{cao2011}. 

During the outburst decay of MAXI J1820+070, the radio luminosity started to decrease after the magnetic field reached the MAD criterion, since both the accretion rate and the relevant magnetic field fed into the ADAF decreased with time \citep{you2023}, which led to the 8 days delay between the radio and Comptonization luminosity. The different cross-correlations between radio and Comptonization in MAXI J1348-630 and MAXI J1820+070 implies that the coupling between the ADAF and jet are complex, which may depend on some properties of the different BHXRB systems.


\subsection{Correlations between X-ray Comptonization luminosity and the radio luminosity}
Correlations between the radio and X-ray luminosity (R-X correlation, $L_\mathrm{R}\propto L_\mathrm{X}^{\beta}$) in the BHXRBs were also explored to study the accretion-jet coupling. The X-ray luminosity in a narrow-band 1--10 keV (or 3--9 keV) was usually used in the studies of R-X correlations \citep{corbel2003, russell2006}. \cite{carotenuto2021/505} studied the R-X correlations using the X-ray luminosity in 1--10 keV obtained from \emph{Swift}/XRT. They obtained a slope $\beta = 0.95$ for $L_\mathrm{X} > 6.3 \times 10^{35} \rm erg \, s^{-1}$. Here, we revisit the R-X correlations, using the observations of \emph{Insight}-HXMT. We convert the radio and X-ray comptonization fluxes to the radio and comptonization luminosities, using a distance of 2.2 kpc \citep{chauhan2021} \footnote{Note that the distance to the source was also estimated to be $D \approx 3.4 \pm 0.3$ kpc, based on X-ray detections of a dust-scattering halo \citep{lamer2021}.}, which is consistent with \cite{carotenuto2021/505}. It turns out that we derived the same slope of $\beta = 0.95$, if also using the X-ray luminosity in 1--10 keV. However, this limited-band X-ray luminosity inevitably contains the contributions from the Comptonization by the ADAF/jet, as well as, the reflection by the accretion disk which is illuminated by the Comptonized photons out of the ADAF/jet \citep{you2021}. 

During the rising phase, the X-ray energy spectrum gradually softens over time, with the photon index $\Gamma$ increasing from 1.53 to 2.22. Meanwhile, the reflection component strengthens, as the reflection fraction (defined as the ratio of the coronal intensity illuminating the disk to the coronal intensity reaching the observer) increases from 0.3 to 2.0. As a result, the portion of the Comptonization in the total luminosity decreases over time. The ratio of the 1--10 keV comptonization luminosity to the 1--10 keV total luminosity decreases from 40\% to 5\%.
Thus, the limited-band X-ray luminosity doesn't directly probe the radiation output from the Comptonization source, if the reflection component is comparable to the Comptonzation component. Moreover, the majority of the overall Comptonization in the hard state is emitted around the hard X-ray band \citep{aaz2004,islam2018}, which is thus underestimated if only considering the narrow-band 1--10 keV (or 3--9 keV) flux. 
In \cite{you2023}, the Comptonization luminosity $L_\mathrm{C}$ in the broadband 0.1-100 keV is estimated by the spectral fits, which is utilized in the studies of the radio/X-ray correlation $L_\mathrm{R}\propto L_\mathrm{C}^{\beta}$ for MAXI J1820+070. They found that the correlation during the soft-to-hard state transition apparently deviates from the one in the hard state.

\begin{figure}
    \includegraphics[width=0.46\textwidth]{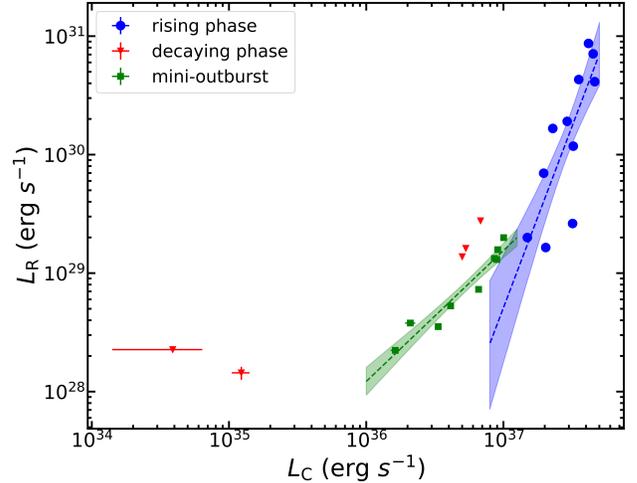}
    \caption{
    The correlation between the radio luminosity at 5 GHz and the X-ray Comptonization luminosity in 0.1-100 keV. The blue, red, and green points represent the data from rising hard state, decaying hard state, and mini outburst, respectively. The linmix algorithm \citep{kelly2007} implemented in Python is used to fit the data points. The dashed lines are the best fits of the radio and X-ray correlation, and the shaded regions indicate uncertainties at the 1-$\sigma$ level.}
    \label{rx}
\end{figure}

In this work, we use the estimated Comptonization luminosity $L_\mathrm{C}$ in 0.1-100 keV and the radio luminosity $L_\mathrm{R}$ from the core location to investigate the R-X correlation in the outburst of MAXI J1348+630. We find that:
\begin{itemize}
    \item   $L_\mathrm{R} \propto L_\mathrm{C}^{3.04 \pm 0.93}$, $\,$ for the rising phase; 
    \item   $L_\mathrm{R} \propto L_\mathrm{C}^{1.11 \pm 0.15}$, $\,$ for the mini-outburst.
\end{itemize}
The corresponding correlations are plotted in Fig. \ref{rx}. Note that, for the rising phase, the time-delay of $\sim$ 3 days between the radio and X-ray is taken into account in the radio-Compton correlation analysis, which results in a strong correlation with Spearman rank correlation coefficients of $\sim 0.8$ and null hypothesis probability of $\sim 0.3\%$. In comparison, if not considering this time delay of $\sim$ 3 days, the correlation becomes worse with Spearman rank correlation coefficients of $\sim 0.6$ and null hypothesis probability (of no correlation) of $\sim 4\%$. For MAXI J1820+070, \cite{you2023} found that $L_\mathrm{C}$ and $L_\mathrm{R}$ follow a correlation $L_\mathrm{R} \propto L_\mathrm{C}^{0.53 \pm 0.06}$ during the rising hard state during which there is no daily delay. However, during the soft-to-hard state transition, the radio emission was discovered to lag the X-ray Comptonization by $\sim$ 8 days. Moreover, if the radio lag of $\sim$ 8 days is not taken into account, the luminosity $L_\mathrm{C}$ and $L_\mathrm{R}$ apparently deviate from the above correlation during the rising hard state. In this work, we revisit the radio/X-ray correlation during the soft-to-hard state transition and the decaying hard state of MAXI J1820+070, taking the radio lag of $\sim$ 8 days into account. It turns out that the slope is $\beta = 0.80$, with Spearman rank correlation coefficients of $\sim 0.98$ and null hypothesis probability less than $\sim 0.1\%$. This is steeper than the derived slope $\beta = 0.53$ during the rising hard state \citep{you2023}.
This highlights the necessity of considering the time delay in the R-X correlation analysis if the time delay is detected.
As for the decaying phase, 
the radio-Compton relation does not follow a single power law (see \autoref{rx}). This has also been found in \cite{carotenuto2021/505} that R-X relation is broken at the 1--10 keV luminosity $L \sim  6.3\times 10^{35} \, \rm erg \, s^{-1}$. Therefore, we cannot derive a monotonic correlation during the decaying phase.

As for the steep R-X correlation with $\beta \sim 3$ derived above, one of the possible scenarios is the jet Doppler boosting \citep[e.g.][]{russell2014}, which requires a low inclination of the source. Up to date, the inclination angle of MAXI J1348-630 is not well constrained, but it was estimated to be low in previous studies, e.g., $\lesssim 46^{\circ}$ \citep{carotenuto2021/504}, and $\sim 29^{\circ}$ \citep{carotenuto2022a}. The alternative cause of such a steep correlation is attributed to the outer thin disk in the truncation scenario being radiation pressure dominated (Y.H. Jiang et al. 2024, in preparation). In this case, the field generated in the thin disk will depend on the accretion rate as $B \sim \dot{m}$. In comparison, if the thin disk is gas press dominated, the field generated will depend on the accretion rate as $B \sim \dot{m}^{3/5}$. Given that the field in the outer disk is dragged through ADAF towards BH, this makes the radio luminosity (jet power) vary more sensitively with the accretion rate, and finally leads to a steeper radio/X-ray correlation. 

As for the mini-outburst, the R-X correlation slope of $\sim 1.1$ is much flatter than the one for the rising phase of the main outburst. Furthermore, there is most likely no delay between the radio and X-ray Compotonization. These indicate an intrinsic difference in the accretion-jet coupling between the main outburst and the mini-outburst, which will be studied in future work.


\section{Acknowledgements}

We acknowledge the referee for valuable comments. We thank F. Carotenuto and S. Corbel for their discussions and help with the data of MAXI J1348-630. B.Y. is supported by Natural Science Foundation of China (NSFC) grants 12322307, 12361131579, and 12273026; by the National Program on Key Research and Development Project 2021YFA0718500; by the Natural Science Foundation of Hubei Province of China 2022CFB167; by the Fundamental Research Funds for the Central Universities 2042024kf1033; by Xiaomi Foundation / Xiaomi Young Talents Program.
X.C. is supported by the NSFC grants 12073023, 12233007, and 12347103; by the science research grants from the China Manned Space Project with No. CMS-CSST- 2021-A06; by the fundamental research fund for Chinese central universities (Zhejiang University).
Z.Y. is supported by the NSFC grants U1838203, U1938114 and 12373049; the Youth Innovation Promotion Association of CAS id 2020265. AAZ acknowledges support from the Polish National Science Center under the grants 2019/35/B/ST9/03944, 2023/48/Q/ST9/00138, and from the Copernicus Academy under the grant CBMK/01/24.

\clearpage

\bibliography{reference}{}
\bibliographystyle{aasjournal} 

\end{document}